\documentclass{nature}
\usepackage{graphicx}
\usepackage{epsf}
\usepackage{lscape}

\title{WASP-12b as a prolate, inflated and disrupting planet from
tidal dissipation}

\author{Shu-lin Li$^{1,2}$, N. Miller$^{3}$,
Douglas N. C. Lin$^{1,3,*}$, Jonathan J. Fortney$^{3}$}

\begin{document}

\maketitle

\begin{affiliations}
\item Kavli Institute for Astronomy and Astrophysics,
Peking University Beijing,100871, China
\item Department of Astronomy, Peking University, Beijing 100871, China
\item Department of Astronomy and Astrophysics, University of California,
Santa Cruz 95064, USA
\end{affiliations}

\begin{abstract}
The class of exotic Jupiter-mass planets that orbit very close to
their parent stars were not explicitly expected before their
discovery$^1$. The recently found$^2$ transiting planet WASP-12b has a
mass $M_p = $ 1.4($\pm 0.1$) Jupiter masses ($M_J$), a mean orbital
distance of only
3.1 stellar radii (meaning it is subject to intense tidal forces), and
a period of 1.1 days. Its radius 1.79($\pm 0.09$) $R_J$ is
unexpectedly large and its orbital eccentricity 0.049($\pm 0.015$) is
even more surprising as such close orbits are in general quickly
circularized. Here we report an analysis of its properties, which
reveals that the planet is losing mass to its host star at a rate
$\sim 10^{-7} M_J$ yr$^{-1}$. 
The planet's surface is distorted by the star's gravity and the light curve produced by its prolate shape will differ by about ten per cent from that of a spherical planet.
We conclude that dissipation of the star's tidal perturbation in the planet's convective envelope provides the energy source for its large volume.
We predict up to 10mJy CO band-head (2.292 $\mu$m)
emission from a tenuous disk around the host star, made up of tidally
stripped planetary gas. It may also contain a detectable resonant
super-Earth, as a hypothetical perturber that continually stirs up
WASP-12b's eccentricity.
\end{abstract}

Gas giant planets contract as they age and cool. Theoretical
models$^{3,4}$ predict an upper radius limit $\simeq 1.2 R_J$ for
mature Jupiter-like gas giants. The radii ($R_p$) of $\sim 60$
short-period (mostly $P<5$ days) gas giants have been measured from
transit light curves. In the mass range $M_p \simeq (1 \pm 0.5) M_J$,
most planets' $R_p \simeq (1.2 \pm 0.2) R_J$. However, several
planets have observed $R_p \sim 1.5-1.8 R_J$. The rate of heat loss from
these planets increases with their radius. The over-sized WASP-12b has
an intrinsic radiative luminosity of $L_p \sim 1-2 \times 10^{28}$erg
s$^{-1}$. If gravity provides the only source with which to replenish this heat
loss, WASP-12b would contract$^{3,5}$ significantly over 100 million years.
The small orbital separation from its host star means that WASP-12b is one of the most intensely heated planets known. Stellar photons deposit energy
onto a planet's day side, and reduce the internal heat loss$^{6}$, but
the absorbed stellar irradiation is efficiently re-radiated along with
the heat flux from a planet's interior on its night side$^{7}$. 
Although surface evaporation enlarges a planet’s tenuous atmosphere$^{8}$,
it cannot alone significantly modify its internal structure and evolution$^{9}$.
 
To account for the inflated sizes of gas giants that are gigayears old, we
consider additional heating mechanisms. Given its measured
eccentricity $e_p \simeq 0.05$ and its proximity to its host star, one
potential suitable energy source for WASP-12b (and other close-in
planets) is the dissipation of the stellar tidal disturbance$^{10}$
well below its photosphere. A planet's tidal heating rate ${\dot E}_t
\sim e^2 G M_p M_\ast / a \tau_e$ (where $G$ is the gravitational constant and $a$ is the semi-major axis.) is determined by its circularization
time scale $\tau_e$ (where subscript $e$ indicates eccentricity damping), which is proportional to its tidal quality
factor$^{11}$, $Q_p ^\prime$. (Here M* is the stellar mass.) We note that many 
long-period planets have
elongated orbits ($0< e_p <1$) whereas those of short-period gas
giants are mostly circular (within the detection limit, $e_p
<0.02$). If this transition is due to their own life-long tidal
dissipation, their inferred $Q_p ^\prime \sim 10^{6} to 10^{7}$.                              

For its orbit, WASP-12b's $\tau_e \sim 0.033 Q_p^\prime (X_R/R_p)^5$
years, where $X_R \equiv (M_p/3 M_\ast)^{1/3} a$ is the distance from the
planet's core to its inner Lagrangian ($L_1$) point (that is, the Roche radius).
If WASP-12b's
$Q_p^\prime \sim 10^7$, then its eccentricity would be damped within ten million years
and the associated tidal heating rate would balance its current
intrinsic radiative luminosity (that is, ${\dot E}_t \sim L_p$).
Depending on their internal structure and orbital periods, the
magnitude of $Q_p ^\prime$ of gas giants and their host stars are
thought to vary over a large magnitude range$^{12,13}$ (see below). If
WASP-12b's $Q_p^\prime$ is comparable to that inferred for
Jupiter$^{14}$ ($\sim 10^6$), its tidal heating would over-compensate
beyond its intrinsic luminosity and would drive ongoing envelope
expansion. Once the planet's atmosphere filled its Roche lobe, mass
would be lost at a rate of ${\dot M}_p \sim {\dot E}_t R_p/ 2 G M_p$.

To show that WASP-12b is currently losing mass, we
analyse the amount of stellar light blocked during its primary
transit by its cross-section in the (Y-Z) direction normal to the
line (X) joining it and its host star (see Fig. 1). At the
planet's occultation radius $R_p$ and maximum Y distance ($Y_R = 2
X_R/3$) on the planet's Roche lobe, we estimate the atmosphere
density to be $\rho (R_p) \sim 5 \times 10^{-8}$g cm$^{-3}$ and
$\rho (Y_R) \sim 3 \times 10^{-12}$g cm$^{-3}$ respectively (see
Supplementary Information). An interesting consequence of the
planet's prolate shape is that, as a function of orbital phase, it
will scatter incident stellar flux, and emit its own thermal
radiation, differently than a spherical planet would. 
We estimate that this effect leads to 10
Optical and mid-infrared
orbit-modulated flux has been detected from several planets$^{15}$
and WASP-12b is a promising candidate for detection as well.

Near L$_1$, gas falls towards the host star through a nozzle with a
radius $\Delta Y \sim 0.22 X_R$ (see Supplementary Information).
The resultant pressure gradient on the Roche lobe drives gas from
other regions to flow towards L$_1$. This advection along
equipotential surfaces offsets the hydrostatic equilibrium normal
to these surfaces. Consequently, the planet's lower atmosphere expands (see
Fig. 1) at the sound speed ($c_{sound}$) across the Roche lobe (which
has a total surface area $A_R \sim 4 \pi X_R Y_R$) with a mass
flux $\dot M_{\rm observed} = \rho(Y_R) c_{sound} A_R \sim 10^{-7} M_J {\rm
yr}^{-1}$. If WASP-12b's $Q_p ^\prime \sim 10^6$, its $\dot M_p
\sim \dot M_{\rm observed} $ and its atmosphere loss would be
continually replenished by the tidal inflation of its envelope.

Gas slightly beyond the nozzle accelerates towards the star and attains
a free fall speed ($v_f$) and a local density $\rho(L_1) \simeq
\rho(Y_R) (c_{sound} A_R / v_f \pi \Delta Y^2 ) \sim \rho (Y_R)$. Since $
\rho(L_1) < < \rho (R_p)$, this stream remains optically thin. It does
not directly strike WASP-12 but forms a disk which is tidally
truncated$^{16}$ by WASP-12b at $\sim 0.7 a = 3 R_\odot$. Provided
WASP-12's magnetic field has a subsolar strength, the disk extends to
the stellar surface and intercepts a significant fraction ($\sim 0.1$)
of the stellar visual luminosity $^{17}$. Viscous stress also
generates heat at a rate $\sim 10^{-2} L_\odot$ while it transfers
mass and angular momentum $^{18}$ (with an efficiency factor $\alpha
\sim 10^{-2}-10^{-3})$ to produce a disk surface density $\Sigma \sim
\alpha^{-1}$ g cm$^{-2}$.

The disk's warm ($\sim 3000-4000$K) midplane is sandwiched by heated surface
layers which emit continuum and line radiation, respectively. At these
temperatures, CO molecules are preserved and their 2-0 band-head emission at
$\sim$2.292 $\mu$m ($1-10$ mJy at WASP-12's 267 pc distance from the Sun) is
comparable to that of their continuum$^{19}$. Although the expected flux is 1-2
orders of magnitude below that found around some known young stellar objects,
this disk signature may be marginally observable with existing near
infrared spectrometers. Pollution of a few Earth masses of metals may also
enhance the [Fe/H] of star's low-mass ($<10^{-3} M_\odot$) convection zone
depending on the efficiency of a double diffusive instability$^{20}$.

As its tidal debris flows through the disk onto the surface of its host star,
WASP-12b gains orbital angular momentum. Its orbit also exchanges angular
momentum with its host star at a rate which is regulated by the dissipative
efficiency of its tidal perturbation within the stellar envelope$^{21, 22}$.
WASP-12b would spiral toward its host star on a time scale $\tau_a \sim 10$Myr
if the quality factor ($Q_\ast ^\prime$) of its slowly spinning host star is
comparable to that ($\sim 10^6$) estimated from the circularization periods of
binary stars in open clusters $^{23}$.

The low probability of catching a brief glimpse of WASP-12b's rapid
orbital evolution and mass loss can be circumvented with variable
$Q_\ast ^\prime$-values$^{13}$. We suggest that during WASP-12's main
sequence evolution, its $Q_\ast ^\prime$ may have been larger than
$10^8$ such that its planet's orbit did not evolve significantly.
But, as it evolves onto its subgiant track, WASP-12's intrinsic
oscillation frequencies are modified with the expansion of its radius,
deepening of its convective zone, and slowdown of its spin. When these
frequencies are tuned to match with those of the planet's tidal
perturbation, the star's dynamical response is greatly amplified. In
these resonant episodes, $Q_\ast ^\prime$ may decrease$^{13}$ below
$\sim 10^7$, leading to the present epoch of intense star-planet tidal
interaction and dissipation.

Our result for the mass loss from WASP-12b is robust, whereas the
tidal heating scenario depends on the assumed $Q^\prime _p < 10^7$
today. But, the preservation of WASP-12b's finite eccentricity
implies that its $Q_p ^\prime > 10^9$ for most of its life. It is
possible that during its recent accelerated orbital decay (due to a
decline in $Q_\ast ^\prime$), the evolving stellar perturbation
frequency and strength may have intensified the planet's tidal response
and dissipation over some ranges of its orbital period. The subsequent
expansion of the heated envelope leads to further changes in the
planet's oscillation frequencies and enlarges the range of efficient
tidal dissipation. We estimate an effective $Q_p^\prime (\sim
10^{6-7})$ for the present-day WASP-12b which is comparable to that
needed to maintain its inflated size.

It is also possible that WASP-12b's eccentricity is continually
excited by one or more super-Earths with periods $P_2 < P$ and masses
$M_2 < M_p (P_2/P)^{13/3}$. Their convergent paths would lead to
resonant capture$^{24}$ followed by lock-step orbital decay and $e$
growth to an equilibrium$^{25}$.
Episodic declines of $Q^\prime _\ast$ and $Q^\prime _p$ (over a few
Myr) would excite WASP-12b's equilibrium eccentricity and inflate its radius to their
observed values. Such a hypothetical planet would be embedded$^{16}$
in the circumstellar disk and the orbital migration induced by its
tidal interaction with the disk$^{26}$ (on time scales
$> 10^8$ yr) would not affect its
resonant interaction with WASP-12b. The sensitivity of its
radial velocity detection would need to be $< 1$ms$^{-1}$ and its transit observation would need to be 10$^{-1}$ magnitude.

Our analysis can be used to constrain the efficiency of tidal
dissipation of other close-in planets such as the recently
discovered$^{22}$ massive (10$M_J$), short-period (0.94 d) WASP-18b.
If it and its host star have similar $Q^\prime$ values ($\sim 10^6$),
WASP-18b's tidal dissipation (at a rate $\sim 10^{30}$erg s$^{-1}$)
would maintain the synchronization of its spin$^{10}$ and induce mass
loss through Roche-lobe overflow during the decay of its orbit
(induced by the star's tidal dissipation) on a time scale of $\sim 1$
Myr. However, WASP-18b is observed to have a normal radius ($1.1
R_J$) for massive gas giants which places an upper limit ($<
10^{28}$erg s$^{-1}$) on the planetary dissipation rate$^{5}$ which
implies either $Q_\ast ^\prime > 10^7$ or $Q_p ^\prime > 10^{10}$ (or
both). This constraint include the possibility for WASP-18b to be
preserved (with $Q_\ast ^\prime > 10^9$) until the end of its host
star's main sequence evolution.

\begin{figure}
\includegraphics[angle=0]{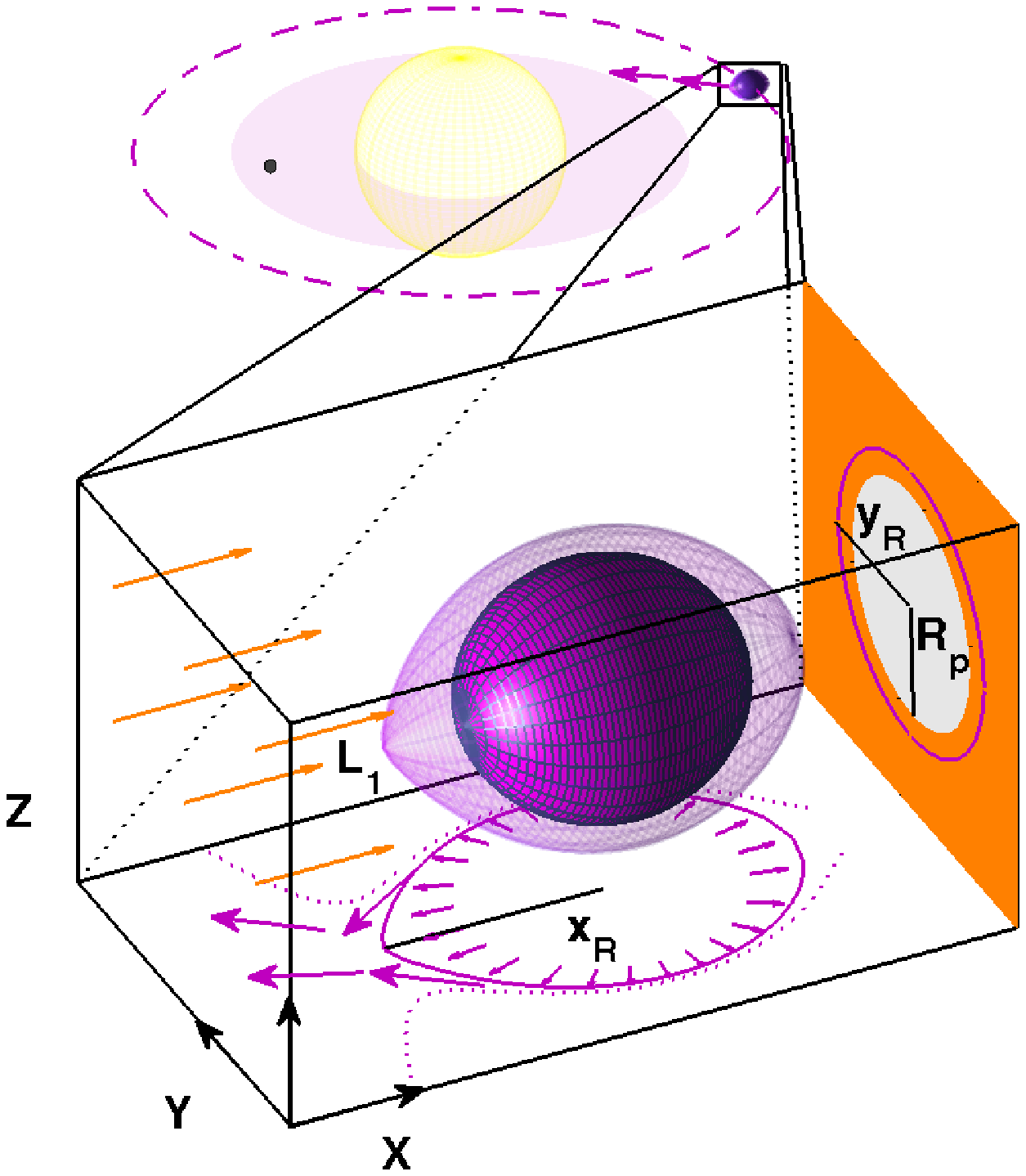}
\caption{WASP-12b's surfaces. 
The inner opaque purple surface (We note its prolate shape) contains the planet's envelope, which contributes to the eclipse of the stellar light, as inferred from transit observations. The stellar photons are represented by the orange arrows.
The planet's outer transparent surface is the L$_1$ equal
potential surface and its projection onto the orbital planet is contained in the solid purple line. The equipotential surfaces are computed
assuming that the planet is a point mass on a nearly circular
orbit. WASP-12b has a thin atmosphere that extends between
these two surfaces; some of this atmosphere flows through
the inner Lagrange point and eventually onto WASP-12. The purple dotted lines indicate the projection of the equipotential surface which channels the flow (purple arrows) from the planet's expanding envelope. The presence of a tenuous (light purple) disk with a hypothetical embedded planet (black dot) around WASP-12 (light yellow sphere) is illustrated at the top. The planet's orbit is traced by the purple dashed line.
All variables are defined in the text.}
\end{figure}





\begin{thebibliography}{1}

\bibitem{dummy1} Mayor, M. \& Queloz, D. A Jupiter-Mass Companion to a
Solar-Type Star. {\it Astrophys. J.} {\bf 378}, 355-359 (1995).

\bibitem{dummy2} Hebb, L. {\it et al.}
WASP-12 b: the hottest transiting extra solar-planet yet discovered.
{\it Astrophys. J.} {\bf 693}, 1920-1928 (2009).

\bibitem{dummy3} Bodenheimer, P., Laughlin, G. \& Lin, D. N. C. On the Radii
of Extrasolar Giant Planets. {\it Astrophys. J.} {\bf 592}, 555-563 (2003).

\bibitem{dummy4} Fortney, J. J., Marley, M. S. \& Barnes, J. W. Planetary
Radii Across Five Orders of Magnitude in Mass and Stellar Insolation:
Application to Transits. {\it Astrophys. J.} {\bf 659}, 1661-1672 (2007).

\bibitem{dummy5} Bodenheimer, P., Lin, D. N. C. \& Mardling, R. A. On
the Tidal Inflation of Short-Period Extrasolar Planets. {\it
Astrophys. J.} {\bf 548}, 466-472 (2001).

\bibitem{dummy6} Guillot, T., Burrow, A., Hubbard, W. B., Lunine, J. I. \&
Saumon, D. Giant Planets at Small Orbital Distances. {\it Astrophys.
J.}, {\bf 459}, L35-L38 (1996).

\bibitem{dummy7} Dobbs-Dixon, I. \& Lin, D. N. C. Atmospheric Dynamics of
Short-Period Extrasolar Gas Giant Planets. I. Dependence of Nightside
Temperature on Opacity. {\it Astrophys. J.}, {\bf 673}, 513-525 (2008).

\bibitem{dummy8} Garcia Munoz, A., Physical and chemical aeronomy of
HD 209458b. {\it P\&SS} {\bf 55}, 1426-1455 (2007).

\bibitem{dummy9} Hubbard, W. B., Hattori, M. F., Burrows, A., Hubeny,
I. \& Sudarksy, D. Effects of mass loss for highly-irradiated giant
planets. {\it Icarus} {\bf 187}, 358-364 (2007).

\bibitem{dummy10} Dobbs-Dixon, I., Lin, D. N. C. \& Mardling, R. A.
Spin-Orbit Evolution of Short-period Planets. {\it Astrophys. J.}
{\bf 610}, 464-476 (2004).

\bibitem{dummy11} Goldreich, P. \& Soter, S. Q in the Solar System.
{\it Icarus}, {\bf 5}, 375-389 (1966).

\bibitem{dummy12} Ogilvie, G. I. \& Lin, D. N. C. Tidal Dissipation in
Rotating Giant Planets. {\it ApJ} {\bf 610}, 477-509 (2004).

\bibitem{dummy13} Ogilvie, G. I. \& Lin, D. N. C. Tidal Dissipation in Rotating
Solar-Type Stars. {\it Astrophys. J.} {\bf 661}, 1180-1191 (2007).

\bibitem{dummy14} Yoder, C. F. \& Peale, S. J. The Tide of Io. {\it
Icarus} {\bf 47}, 1-35 (1981).

\bibitem{dummy15} Knutson, H. {\it et al.} A map of the day-night contrast of the extrasolar planet HD 189733b. {\it Nature} {\bf 447}, 183-186 (2007).

\bibitem{dummy16} Lin, D. N. C. \& Papaloizou, J. C. B. On the tidal
interaction between protostellar disks and companions. {\it in
Protostars and Planets III} (eds: D. Black and M. Mathews, University
of Arizona Press). 749-835 (1993).

\bibitem{dummy17} Adams, F. C., Shu, F. H. \& Lada, C. J. The disks of T
Tauri stars with flat infrared spectra. {\it Astrophys. J.} {\bf 326},
865-883 (1988).

\bibitem{dummy18} Hartmann, L., Calvet, N., Gullbring, E. \& D'Alessio,
P. Accretion and the Evolution of T Tauri Disks. {\it Astrophys. J.}
{\bf 495}, 385-400 (1998).

\bibitem{dummy19} Najita, J., Carr, J. S., Glassgold, A. E., Shu, F. H. \&
Tokunaga, A. T. Kinematic Diagnostics of Disks around Young Stars: CO Overtone
Emission from WL 16 and 1548C27. {\it Astrophys. J.} {\bf 462}, 919-936 (1996).

\bibitem{dummy20} Vauclair, S. Metallic Fingers and Metallicity Excess in
Exoplanets' Host Stars: The Accretion Hypothesis Revisited.
{\it Astrophys. J.} {\bf 605}, 874-879 (2004).

\bibitem{dummy21} Sasselov, D. D. The New Transiting Planet OGLE-TR-56b:
Orbit and Atmosphere. {\it Astrophys. J.} {\bf 596}, 1327-1331 {2003}.

\bibitem{dummy22} Hellier, C., {\it et al.}
An orbital period of 0.94days for the hot-Jupiter planet WASP-18b.
{\it Nature} {\bf 460}, 1098-1100 (2009).

\bibitem{dummy23} Meibom, S., Mathieu, R. D. \& Stassun, K. G. An
Observational Study of Tidal Synchronization in Solar-Type Binary
Stars in the Open Clusters M35 and M34. {\it Astrophys. J.} {\bf
653}, 621-635 (2006).

\bibitem{dummy24} Peale, S., Cassen, P. \& Reynolds, R. T. Melting Io by
Tidal Dissipation. {\it Science} {\bf 203}, 892-894 (1979).

\bibitem{dummy25} Lin, D. N. C. \& Papaloizou, J. On the structure of
circumbinary accretion disks and the tidal evolution of commensurable
satellites. {\it Mon. Not. R. Astron. Soc.} {\bf 188}, 191-201 (1979).

\bibitem{dummy26} Tanaka, H., Takeuchi, T. \& Ward, W. R.
Three-Dimensional Interaction between a Planet and an Isothermal
Gaseous Disk. I. Corotation and Lindblad Torques and Planet Migration.
{\it Astrophys. J.} {\bf 565}, 1257-1274 (2002).

\end{thebibliography}

\section*{Supplementary Information}
Outside $R_p$, an optically thin atmosphere establishes a quasi
hydrostatic equilibrium which extends to the planet's Roche-lobe
surface. Neglecting small temperature, sound speed ($c_s$), and
opacity ($\kappa_0$) variations, the gas density at a radial location
$R^\prime (> R_p)$ is
\begin{equation}
\rho (R^\prime) = \rho (R_p) {\rm exp} [- \lambda (1 - {R_p / R^\prime})],
\end{equation} 
where $\lambda = G M_p / c_s^2 R_p$. Assuming spherical symmetry, we
find $ \rho(R_p) = [ \kappa_0 f(\theta_1) R_p]^{-1}$ where
\begin{equation}
f(\theta_1) \equiv \int_{-\theta_1} ^{\theta_1} (d \theta / {\rm
cos}^2 \theta) {\rm exp}[ - \lambda (1 -{\rm cos} \theta)],
\end{equation}
is a weak function of $\theta_1 \equiv {\rm cos}^{-1} (Y_R / R_p)$.
WASP-12b's atmospheric temperature at $X_R$ is $\sim
2,500$K for effective planet-wide energy redistribution of absorbed
stellar flux and $\sim 3000$K on the day-side for inefficient thermal
circulation. A self-consistent analysis of transit data$^{27}$ yields
$\kappa\sim 10^{-2}$ cm$^2$ g$^{-1}$ at $R_p$. With corresponding
values of $c_s \sim 5$ km s$^{-1}$, $\lambda =59$, and $f=0.165$ at
pericenter (where $X_R=3.2 R_J$ and $Y_R = 2.1 R_J$), we find $\rho
(R_p) \sim 5 \times 10^{-8}$ g cm$^{-3}$ and $\rho (Y_R) \sim 3 \times
10^{-12}$g cm$^{-3}$ respectively.

On the Roche lobe, gas flows toward L$_1$ and is channeled into a
stream which falls toward the host star with a flux ${\dot
M}_{obs}$. Gas in the stream is confined by the Y opening of the
equipotential surface which contains atmosphere at one unperturbed
scale height $Y_R/ \lambda = Y_T c_s^2 R_p/ G M_p $ outside
$Y_R$. The radius of the nozzle is $\Delta Y = (2/3)^{1/2} c_s P/2 \pi
\sim 0.4 R_p \sim 0.22 X_R$.  Despite the convergent flow, gas density
along in the stream is $\sim \rho (Y_R)$ as its main flow velocity is
accelerated to the supersonic free fall speed.  Since $\rho (Y_R) < <
\rho (R_p)$, this stream is optically thin.

We thank an anonymous referee for pointing out that the mass loss rate
due to the X-ray and EUV driven thermal evaporation of the upper
atmosphere is negligible compared with $\dot M_{\rm obs}$. WASP-12 may
also have a wind with a flux $\dot M_w$ and speed $v_w$ comparable to
those of the Sun ($\sim 10^{-13} M_\odot$ yr$^{-1}$ and $<10^3$ km
s$^{-1}$). The wind's ram pressure ${\dot M_w} v_m/ 4 \pi a^2$ on
WASP-12b's day side could potentially exceed the planet's atmospheric
thermal pressure and perturb the flow toward the night side$^{28}$ but
it is inadequate to push the gas to overcome the gravitational
potential difference$^{29}$ $2 G M_\ast / 3 a$ between $L_1$ and the
outer Lagrangian point at $L_2$. The ongoing expansion of WASP-12b's
envelope would continually increase its atmospheric density until a
mass-flow equilibrium is established with an outflow through the $L_1$
nozzle. Finally, if WASP-12b has a magnetic field larger than a few
Gauss, outflow of a tenuous, highly ionized atmosphere could be
suppressed. However, the accumulation of the expanding envelope would
also quench the magnetic influences in WASP-12b's atmosphere and lead
to an outflow equilibrium.


\begin{addendum}

\item[Supplementary Information] is linked to the online version
of the paper at www.nature.com/nature.

\item[Acknowledgements] This work is supported by the Kavli Foundation which
enabled the principle initiation and development of this work at KIAA-PKU. It
is also supported by NASA, JPL, and NSF.

\item[Author contributions] Shulin Li and D.N.C. Lin constructed
arguments for mass loss and tidal heating of WASP-12b. They also composed the
draft of the paper. N. Miller brought to the attention of the team on
WASP-12b's large radius and designed the illustration. J. Fortney contributed
information on the planet's opacity and improved the presentation of the
manuscript.

\item[Competing Interests] The authors declare that they have no
competing financial interests.

\item[Author Information] Correspondence and requests for materials
should be addressed to Douglas N. C. Lin~(email: lin@ucolick.org),
Department of Astronomy and Astrophysics, University of California,
1156 High Street, Santa Cruz, CA 95064, Fax: 831 459 5265.
\end{addendum}

\end{document}